\documentclass[floatfix,aps,tightenlines,prd,nofootinbib,superscriptaddress]{revtex4}
\usepackage{epsfig}

\usepackage{color}              

\begin{document}
\title{Self force of a static electric charge near a Schwarzschild Star}
\author{Karthik Shankar and Bernard F Whiting}
\affiliation{Department of Physics, University of Florida,
Gainesville, FL 32611-8440, USA}

\begin{abstract}
When a charge is held static near a constant density spherical star, it experiences a self-force $\mathrm{F}_{\mathrm{self}}$, which is significantly different from the force $\mathrm{F}_{\mathrm{self}}^{\mathrm{BH}}$ it would experience when placed near a black hole of the same mass.  In this paper,
an expression for
the self-force (as measured by a locally inertial observer) is given 
for an insulating Schwarzschild star, and the result is explicitly computed for the extreme density case, which has a singularity at its center. The force is found to be repulsive. A similar calculation of the self-force is also performed for a conducting star. This calculation is valid for any static, spherically conducting star, since the result is independent of the interior metric.  When the charge is placed very close to the conducting star, the force is found to be attractive but when the charge is placed beyond a certain distance (2.95M for a conducting star of radius 2.25M), the force is found to be repulsive. When the charge is placed very far from the star (be it conducting or insulating), the charge experiences the same repulsive force it would experience when placed in the spacetime of a black hole with the same mass as the star.
\end{abstract}

\maketitle{}

\section{Introduction }\label{intro}

The electrostatic potential and the fields produced by a static
electric charge in the vicinity of a Schwarzschild black hole have
been discussed in detail in several
papers \cite{copson,linet,wald,hanni}.
These all suppose that an external force holds an electric charge $e$
at rest at a coordinate position $R=b$ in the Schwarzschild
geometry. This external force should balance the electrostatic force
on the charge in addition to the usual attractive gravitational
force (on the mass carrying the charge). When 
there is no other charge anywhere,
the entire electrostatic force on the charge may be regarded as a
self-force.

An explicit expression for the electrostatic potential of such a point
charge is given as a summation over multipole moments in
\cite{wald}. The horizon of the black hole is found to be an
equipotential surface and \cite{hanni} shows the electric lines
of force everywhere outside the horizon. A closed form expression
for a specific potential, $V^{\rm C}$, was given earlier by Copson \cite{copson},
using work of Hadamard \cite{hadamard},
but it corresponds to a solution with non-zero charge on the
black hole. For the potential corresponding to
physically relevant boundary conditions (namely, zero net charge on
the black hole), Linet\cite{linet} observed that a spherically symmetric homogeneous solution had to be added to Copson's potential, yielding a result in accord with \cite{wald}.


By employing a strategy similar to Dirac's\cite{dirac},
of imposing conservation of the stress energy tensor inside a
world-tube surrounding the particle and then limiting the world tube
to the world line of the particle, Smith and Will\cite{Will} 
calculated the external force needed to hold this charge at rest.
After subtracting the requisite gravitational force, the electrostatic
self-force is found to be $F_{\mathrm{self}}^{\mathrm{BH}}=e^{2}M/b^{3}$.\footnote[1]{We use units in which $G=c=1$ throughout.}  It is implicit that
this force is always calculated with respect to a locally inertial
observer at the position of the charge.  The potential of \cite{wald}
or \cite{linet} is essential for understanding this result:  with the potential, $V^{\rm C}$, given by Copson\cite{copson}, no self-force would have been found.

A static electric charge is a singular point source and the potentials discussed so far all satisfy Maxwell's equations with the point particle as source.  This is quite distinct from the behavior of the so-called ``direct'' and ``tail'' fields \cite{dewitt:brehme}, for which the source in Maxwell's equations is not well identified.  By contrast, the radiation potential of Dirac\cite{dirac} satisfies Maxwell's equations with no source.  Other familiar potentials which satisfy Maxwell's equations with the point particle as source include the retarded and advanced potentials, and their symmetric sum.  Ironically, Copson's potential in \cite{copson} is none of these familiar constructs, nor is the potential of \cite{wald, linet}.

Detweiler and Whiting\cite{bernard} demonstrate that, in general, a
potential $V$ due to a point source can be written as sum of two
parts, $V=V^{\rm S}+V^{\rm R}$, where $V^{\rm S}$ is a singular
potential (divergent at the particle) which exerts no self force on
the particle and $V^{\rm R}$ is the regular potential which is
entirely responsible for the self-force. In particular, locally near the source, $V^{\rm S}$ satisfies Maxwell's equations with the point particle as source, while $V^{\rm R}$ is an homogeneous solution, 
without source.
Within a normal neighborhood of the point charge, the singular potential can
be obtained from a variant of Hadamard's form of the Green's
function\cite{hadamard}.  We will need to understand its relation to the potential, $V^{\rm C}$, of Copson, that is, locally.

\subsection{The Singular Potential}

It can be understood that Copson's potential for this static
problem is exactly $V^{\rm S}$ because of the relation between the
construction of $V^{\rm S}$ in \cite{bernard} and Copson's
construction of $V^{\rm C}$ using Hadamard's work. In constructing
the potential we refer to as $V^{\rm C}$, Copson used the unique,
locally-represented, least-singular (hence) elementary solution
for the static problem, as given in a general formalism due to
Hadamard\cite{hadamard}.  By construction, the singularity in
$V^{\rm C}$ at the source is of as low an order as is possible.
In the not necessarily static case, the singular potential $V^S$
is, in a normal neighborhood of the point charge, the unique
locally-determined, similarly least-singular solution of
generalized Hadamard form\cite{bernard}, which has support on and
outside the light-cone (it exerts no force).  In this regard,
$V^{\rm S}$ is unique, due to its lack of (local) support inside
the light cones.  In the static case, when the light-cone
collapses and time derivatives disappear in the equation for the
potential, the domain of support for $V^S$ becomes precisely the
domain in which $V^{\rm C}$ is defined.  Then, $V^{\rm C}$ and
$V^{\rm S}$ solve the same problem under the same conditions, and
hence coincide.


The result that Copson's potential $V^{\rm C}$ is precisely
$V^{\rm S}$, is indeed compatible with the fact that it yields no
self-force. Consequently, the correction to Copson's potential
given in \cite{linet} automatically corresponds to $V^{\rm R}$,
the regular part of the potential, which provides the self force.

\subsection{Overview}

In this paper, we consider replacing the black hole by a constant density star.  The fundamental fact we will use is that, locally, the potential
constructed by Copson $V^{\rm C}$ is exactly the singular piece $V^{\rm
S}$, as long as the metric in the neighbourhood of the charge is the
Schwarzschild metric. Any additional part of the potential at the
position of the charge will contribute to the self force. Our aim is
to calculate the self force of a charge $e$ placed in the vicinity
of a spherical star in two different situations --- depending upon whether
the star is totally conducting or totally insulating. Since the metric outside
the star is the Schwarzschild metric, Copson's potential is indeed
$V^{\rm S}$. To calculate the self force, we follow Linet\cite{linet} by first finding the homogenous solution of Maxwell's equations $\widetilde{V}$, which
should be added to $V^{\rm S}$ (given by Copson) to give the whole potential.

In section \ref{metric}, we briefly discuss the Schwarzschild black hole
metric, and the metric of a constant density (Schwarzschild) star,
which we show is conformally related to a much simpler metric.
In section \ref{staticEM}, we obtain the solutions to Maxwell's equations inside
and outside the star. We start with the singular potential $V^{\rm
S}$ outside the star and continue the solution inside the star's
surface. This reveals that the surface of the star carries a charge
distribution, a result which is not physically acceptable.  We correct this in  section 4, where we impose two different, physically 
interesting, boundary conditions to calculate the homogeneous solution
$\widetilde{V}$, which in turn is responsible for the self force. In
section \ref{insulate}, we consider an insulating star that does not get
electrically polarized. An additional, regular field, arising both inside and outside the star, is necessary to cancel the surface charge induced by $V^{\rm S}$.  It is this additional field which is responsible for the self force on the particle.  We find that the self force
$F_{\mathrm{self}}$, is repulsive and is significantly different
from $F_{\mathrm{self}}^{\mathrm{BH}}$ when the charge is placed
very close to the star. In section \ref{cond}, we consider a conducting
star which has no electric field in the interior, but nevertheless requires an additional field in the exterior to remove the net surface charge which would otherwise be induced on the star. Here, we find that
$F_{\mathrm{self}}$ is attractive when the charge is placed close to
the star ($b<2.95M$) and is repulsive when the charge is placed
farther than $2.95M$. We should interpret the attractive force as a
consequence of electrical polarization of the star and the repulsive
force as due to interaction of the fields with space-time curvature.

\subsection{Discussion}

For calculating the self force of a point source in curved
spacetime, a variety of regularization techniques have been used\cite{dewitt:brehme, barack:ori, quinn:wald, bernard}. In a case somewhat related to ours, with a point source placed near a thin shell, Burko et al.\cite{burko:etal} have explicitly shown how the regularization techniques can be used to calculate the self force. In this paper, we do not perform a regularization to calculate the self force, but we very much
rely on the regularization details in \cite{Will, bernard}.

We have focused on the extreme density case because of its intrinsic interest.
In principle, our approach here can be used to calculate the
self-force on a static electric charge in the vicinity of any
star with a spherically symmetric metric, not just a star with constant density. Generally, the resulting expression for
the self-force would be very complicated and would require numerical
evaluation, which is actually what we resort to in the final step of
evaluation of the self-force in this paper.  It would be interesting to investigate whether such results could have any detectable effect on neutron star physics and observations.

Another question that might come to mind is whether the self-force
calculated for the static charge $F_{\mathrm{self}}$, has any
connection with the self force experienced by a freely falling
charge on which no external force acts. DeWitt and DeWitt
\cite{dewitt} have shown that the self force on a moving charge
consists of two parts, a radiative part, which removes the energy
from the particle and dumps it into the fields and a conservative
part, which pushes the particle off a geodesic without removing
energy from it. For a slow moving charge, they explicitly show that
the conservative part of the self-force is independent of the
particle's velocity. In the context of the charge placed near a
black hole, Wiseman\cite{wiseman} clearly points out that, the self
force $F_{\mathrm{self}}^{\mathrm{BH}}$ experienced  by the static
charge (as calculated by Will and Smith) exactly corresponds to the
conservative part of the self-force experienced by a freely falling
charge. It is straightforward to see that we can adopt the same
interpretation for the self-force calculated in this paper.

\section{The Metric}\label{metric}

Consider a constant density star of total mass $M$ and radius
$R_{s}$. The metric outside the surface of the star is the Schwarzschild black hole metric (see, for example, \cite{MTW}):
\begin{equation}
ds^{2}=-(1-2M/R)\,dT^{\,2}+\frac{dR^{2}}{(1-2M/R)}+R^{2}d\Omega^{2},\qquad R>R_{s},\label{1}
\end{equation}
and the metric inside the (Schwarzschild) star (see also \cite{MTW}) is
\begin{equation}
ds^{2}=-\left[a-\frac{1}{2}\sqrt{1-\alpha
R^{2}}\right]^{2}dT^{\,2}+\frac{dR^{2}}{[1-\alpha
R^{2}]}+R^{2}d\Omega^{2},\qquad R<R_{s},\label{2}
\end{equation}
where $a=\frac{3}{2}\sqrt{1-2M/R_{s}}$ and $\alpha=2M/R_{s}^{3}$. The
quantity $M/R_{s}$ is bounded above by $4/9$.  Hence the range of $a$ is also restricted:
\begin{equation}
0<M/R_{s}\le 4/9\quad\Rightarrow\quad 3/2>a\ge1/2.
\end{equation}
For the extreme density
case, when $M/R_{s}=4/9$, we have $a=1/2$ and the center, $R=0$,
develops a singularity (the component, $G_{RR}$, of the Einstein tensor
blows up).

The Weyl tensor for the metric in (\ref{2}) evaluates to zero, which means that
the metric is conformally flat. This interior metric can be written in the following simpler form:
\begin{equation}
ds^{2}=\frac{4}{\left(1+\alpha
r^{2}\right)^{2}}\left[-(\beta^{2}+\alpha r^{2})^{2}\,
dt^{2}+dr^{2}+r^{2}d\Omega^{2}\right],\label{3}
\end{equation}
where $\beta^{2}=(2a-1)/(2a+1)$ and the new coordinates are defined as:
\begin{equation}
t=T(2a+1)/4,\quad {\rm and}\quad r=R/(1+\sqrt{1-\alpha R^{2}}).\label{3b}
\end{equation}
The coordinate
$r$ is a monotonic function of $R$, and $r$ ranges from $0$ to some maximum value, $r_{s}$ at the surface of the star. We
have the following expressions for $r_{s}$ and $\beta^{2}$:
\begin{equation}
r_{s}=\frac{R_{s}}{1+\sqrt{1-\frac{2M}{R_{s}}}},\qquad \beta^{2}=\frac{1-2\alpha r_{s}^{2}}{1-\alpha r_{s}^{2}}.
\end{equation}
For each constant density star, $\beta^{2}$ is a constant, and it can take values which range between $0$ and ${1}/{2}$.
For the extreme density case, we have $\beta=0$, and
$r_{s}={3R_{s}}/{4}={27M}/{16}$, in which case the interior
metric takes the following form:
\begin{equation}
ds^{2}=\frac{4}{\left(1+\alpha
r^{2}\right)^{2}}\left[-\alpha^{2}r^{4}\,
dt^{2}+dr^{2}+r^{2}d\Omega^{2}\right].\label{4}
\end{equation}

There exist coordinate transformations which make (\ref{3}) manifestly conformally flat\cite{conf:cords}.  Except possibly for $\beta=0$, those coordinate transformations mix the spatial and temporal coordinates. Our focus is restricted to electrostatics in this paper.
Since Maxwell's equations are
conformally invariant, it will be sufficient to work with the conformal metric evident in (\ref{3}), obtained by ignoring the conformal factor
$4/\left(1+\alpha r^{2}\right)^{2}$.  Thus we will use the metric:
\begin{equation}
d\tilde{s}^{2}=\left[-(\beta^{2}+\alpha r^{2})^{2}\,
dt^{2}+dr^{2}+r^{2}d\Omega^{2}\right],\label{3a}
\end{equation}
when solving 
for the electrostatic potential in the following section.

\section{Electrostatics }\label{staticEM}

Maxwell's equations, which govern the theory of electrodynamics,
can be written as
\begin{equation}
\nabla_{\mu}F^{\mu\nu}=4\pi J^{\nu},
\end{equation}
where $F_{\mu\nu}=\partial_{\mu}A_{\nu}-\partial_{\nu}A_{\mu}$. In terms
of the coordinate derivatives, they can be presented in the following explicit form:
\begin{equation}
\partial_{\mu}(\sqrt{-g}g^{\mu\rho}g^{\nu\sigma}F_{\rho\sigma})=4\pi\sqrt{-g}\, J^{\nu}\label{conf}.
\end{equation}
It is evident that 
Maxwell's equations are conformally
invariant: 
$F_{\mu\nu}$ remains invariant under a
conformal transformation of the metric provided the source term in
(\ref{conf}) is defined so that it, too, is invariant under such a
transformation.\footnote[2]{With $F_{\mu\nu}$ remaining invariant under a conformal transformation, $A_{\nu}$ is also taken to be invariant.ÊÊThen, in terms of $A_{\nu}$, the invariance of the other of Maxwell's equations, $\nabla_{[\alpha}F_{\mu\nu]}=0$, under a conformal transformation, follows trivially from the identity, ${\rm R}^{\sigma}_{(\alpha\mu\nu)}=0$, for the Riemann tensor, which holds in any metric.}
For electrostatics in a static geometry,
Maxwell's equations in the Lorentz gauge ($\nabla_{\mu}A^{\mu}=0$) can be reduced to a single equation:
\begin{equation}
\partial_{i}(\sqrt{-g}g^{ij}g^{00}\partial_{j}A_{0})=4\pi\sqrt{-g}\, J^{0}.\label{Lorentz}
\end{equation}

In this paper, we consider a charge, $e$, on the $z$-axis at the coordinate position
$R=b$, located outside a static, spherical star.
The resulting potential $A_{0}$, which henceforth we denote as
$V(r,\theta)$, will be axially symmetric about the $z$-axis.

\subsection{Outside the star}

To find the potential outside the star, we will solve
Maxwell's equations, which have been reduced to the single equation in (\ref{Lorentz}).
In a static, spherically symmetric 
geometry, 
this takes the form:
\begin{equation}
\partial_{r}(\sqrt{-g}g^{rr}g^{00}\partial_{r}V)+\partial_{\theta}(\sqrt{-g}g^{\theta\theta}g^{00}\partial_{\theta}V)=4\pi\sqrt{-g}\, J^{0},\label{5}
\end{equation}
where
\begin{eqnarray} 2\pi\sqrt{-g}J^{0} &=
e\delta(R-b)\delta(\theta)=e\delta(R-b)\,\sin\theta\delta(1-\cos\theta)\nonumber\\
&=e\delta(R-b)\,\sin\theta{ \sum_{\ell}\left[
(l+\frac{1}{2})P_{\ell}(\cos\theta)\right]}.
\end{eqnarray}
Using the metric in (\ref{1}), equation (\ref{5}) can be written as
\begin{eqnarray}
-\sin\theta\left[\frac{\partial}{\partial R}\left(R^{2}\frac{\partial V}{\partial R}\right)+\frac{1}{(1-2M/R)}\frac{1}{\sin\theta}\frac{\partial}{\partial\theta}\left(\sin\theta\frac{\partial V}{\partial\theta}\right)\right]\nonumber\\
=2e\delta(R-b)\,\sin\theta{
\sum_{\ell}{ {
(\ell+\frac{1}{2})P_{\ell}(\cos\theta)}}}.\label{6}
\end{eqnarray}
We can decompose the solution in terms of Legendre functions as
\begin{equation}
V(R,\theta)=\sum_{\ell}\left[V_{\ell}(R)P_{\ell}(\cos\theta)\right],\label{Legendre}
\end{equation}
and hence simplify (\ref{6}) to get
\begin{equation}
-\frac{\partial}{\partial R}\left(R^{2}\frac{\partial V_{\ell}}{\partial
R}\right)+\frac{\ell(\ell+1)V_{\ell}}{(1-2M/R)}=e(2\ell+1)\delta(R-b).\label{7}
\end{equation}
Copson\cite{copson} constructed a particular solution,
$V^{\rm C}(R,\theta)$, to these equations directly from Hadamard's
elementary solution\cite{hadamard}. Even though the solution he
constructed was in the context of the charge in the presence of a
Schwarzschild black hole, his solution will still be a particular solution
for the case where the black hole is replaced by a spherical star,
because the exterior metric of the spherical star is the
same as the black hole metric. Copson gives
\begin{equation}
 V^{\rm C}(R,\theta)=\frac{e}{bR}\frac{(b-M)(R-M)-M^{2}\cos\theta}{\sqrt{(R-M)^{2}-2(R-M)(b-M)\cos\theta+(b-M)^{2}-M^{2}\sin^{2}\theta}}.\label{8}
\end{equation}
A calculation similar to that in \cite{Will} explicitly shows
that this potential does not contribute to the electrostatic self
force.  
It has been known for some time \cite{bernard}
that the field produced by any point source can be decomposed into two
pieces, a locally-determined, singular field which does not affect the particle's state of motion
and a regular field (which is a solution to the homogeneous field
equations in the neighbourhood of the particle) that does; that is,
$V=V^{\rm S}+V^{\rm R}$. The singular potential
$V^{\rm S}$ has nothing to do with the self-force. It is the
regular potential $V^{\rm R}$ which is entirely responsible for the
self-force.
It was explained in the introduction
that $V^{\rm S}(R,\theta)=V^{\rm C}(R,\theta)$.

The singular field $V^{\rm S}$ is an entirely local construct. The
boundary conditions and external sources play a part only in
determining the regular field $V^{\rm R}$. The regular field interacts with the source and makes it experience a force.
It is crucial in determining the path the charge
actually follows. The external field (due to other sources) is in
general a part of the regular field. Thus, the force experienced by
the point charge is a sum of the force due to external field and the
self-force. In the absence of any external sources, the force
due to the regular field alone corresponds entirely to the self-force.

We can decompose the singular potential (\ref{8}) in terms of Legendre
functions as in (\ref{Legendre}): $V^{\rm
S}(R,\theta)=\sum_{\ell}\left[V_{\ell}^{\rm
S}(R)P_{\ell}(\cos\theta)\right]$.
Using the results of \cite{wald}, 
the multipole moments corresponding to this
potential are given by:

\begin{equation}
V_{\ell}^{\rm S}(R)=e\left\{
\begin{array}{cc}
g_{\ell}(b)f_{\ell}(R)-\frac{M}{Rb}\delta_{\ell,0}   & \text{for   }  R>b>R_s   \\
\vbox{\vskip0.2truein}
f_{\ell}(b)g_{\ell}(R)-\frac{M}{Rb}\delta_{\ell,0}   & \text{for   }  R_s<R<b   \\
\end{array}
\right.
\label{9}
\end{equation}
where
\begin{equation}
f_{\ell}(R)=-\frac{(2\ell+1)!}{2^{\ell}(\ell+1)!\, \ell!\,
M^{\ell+1}}(R-2M)\frac{d}{dR}Q_{\ell}(\frac{R}{M}-1), \quad{\rm and}\label{11}
\end{equation}
\begin{equation}
g_{\ell}(R)=\left\{
\begin{array}{cc}
\frac{2^{\ell}\ell!(\ell-1)!M^{\ell}}{(2\ell)!}(R-2M)\frac{d}{dR}P_{\ell}(\frac{R}{M}-1) &\text{for  } \ell\neq0  \\
1 &\text{for  } \ell=0  \\
\end{array}
\right.
\label{10}
\end{equation}
are the two linearly independent
homogeneous solutions to (\ref{7}).
Note the $\ell\!=\!0$ results: $f_{0}(R)\!=\!1/R$ and $g_{0}(R)\!=\!1$. Otherwise,  as $R\!\rightarrow\!\infty$, 
$f_{\ell}(R)\!\rightarrow\!1/R^{(\ell+1)}$ and $g_{\ell}(R)\!\rightarrow\!1\!\times\! R^{\ell}$.


In the next section, we will calculate the regular field
$V^{\rm R}$ that should be added to the singular field $V^{\rm S}$
in the vicinity of the charge, so as to satisfy the appropriate
boundary conditions.  Before that, we solve for the continuation of $V^{\rm S}$ inside the spherical star.

\subsection{Inside the star}\label{inside}

We assume throughout that the spherical star has a constant density
as discussed in section \ref{metric}. Since
Maxwell's equations in four dimensions are conformally invariant (as
long as the source is conformally invariant), the solution inside
the star can be obtained by using a conformally related metric and
a conformally invariant source in Maxwell's equations. Thus,
we will use the conformal metric, (\ref{3a}), when we solve Maxwell's equations inside the star. The interior potential $V(r,\theta)$, can be found from Maxwell's
equations by using the interior metric (\ref{3a}) in (\ref{5}):
\begin{equation}
 -\sin\theta\left[\frac{\partial}{\partial
r}\left(\frac{r^{2}}{(\beta^{2}+\alpha r^{2})}\frac{\partial
V}{\partial r}\right)+\frac{1}{(\beta^{2}+\alpha
r^{2})}\frac{1}{\sin\theta}\frac{\partial}{\partial\theta}\left(\sin\theta\frac{\partial
V}{\partial\theta}\right)\right]=4\pi
J^{0}\sqrt{-g}.\label{12}
\end{equation}
We assume that there is no charge inside the surface of the
star. Then, $J^{0}\sqrt{-g}$ is the charge density on the star,
which can be expressed as
$\sigma(\theta)\sin\theta\delta(r-r_{s})$, where $\sigma(\theta)$
is the surface charge density.

By axial symmetry, we can decompose the solution  as
$V(r,\theta)=\sum_{\ell}V_{\ell}(r)P_{\ell}(\cos\theta)$. We can
also decompose the charge density as
$\sigma(\theta)=\Sigma_{\ell}\sigma^{\ell}P_{\ell}(\cos\theta)$.
Now, (\ref{12}) reduces to
\begin{equation}
\left[-\frac{\partial}{\partial
r}\left(\frac{r^{2}}{(\beta^{2}+\alpha r^{2})}\frac{\partial
V_{\ell}}{\partial r}\right)+\frac{\ell(\ell+1)}{(\beta^{2}+\alpha
r^{2})}V_{\ell}\right]=\sigma^{\ell}\delta(r-r_{s}).\label{12b}
\end{equation}
The homogeneous solutions to (\ref{12b}) are
obtained in terms of
hypergeometric functions $F([a,b],[c],x)$\cite{Arfken}. For each
$\ell$, there are two linearly independent solutions for $V_{\ell}(r)$,
namely $h_{\ell}(r)$, $k_{\ell}(r)$:

\begin{equation}
h_{\ell}(r)=\left\{
\begin{array}{cc}
\left(-\frac{r}{\beta^{2}}\right)^{\ell}\left(1+\frac{\alpha}{\beta^{2}}r^{2}\right)^{2}\, F\left(\left[\frac{3}{2},2+\ell\right],\left[\ell+\frac{3}{2}\right],-\frac{\alpha}{\beta^{2}}r^{2}\right) &\text{for  } \beta\neq0\\
\vbox{\vskip0.2truein}
(\alpha r)^{-\ell} &\text{ for  } \beta=0,
\end{array}
\right.
\label{13a}
\end{equation}
and
\begin{equation}
k_{\ell}(r)=\left\{
\begin{array}{cc}
\left(-\frac{\beta^{2}}{r}\right)^{\ell+1}\left(1+\frac{\alpha}{\beta^{2}}r^{2}\right)^{2}\, F\left(\left[\frac{3}{2},1-\ell\right],\left[\frac{1}{2}-\ell\right],-\frac{\alpha}{\beta^{2}}r^{2}\right) &\text{for  } \beta\neq0 \\
\vbox{\vskip0.2truein}
(\alpha r)^{\ell+1} &\text{ for  } \beta=0.
\end{array}
\right.
\label{13b}
\end{equation}
Note that $h_{0}(r)=1$, is a constant. When $\beta\neq0$, we see that, for
small values of $r$, $h_{\ell}(r)\,\sim\, r^{\ell}$ and
$k_{\ell}(r)\,\sim\, r^{-\ell-1}$. Since, for $\beta\neq 0$, we want $V_{\ell}$ to be
well behaved at the origin, it should take the form of
$h_{\ell}{(r)}$.  Away from $r=0$, $h_{\ell}{(r)}$ is well behaved in the limit $\beta\rightarrow0$, so we propose that $V_{\ell}$ should also take the form of
$h_{\ell}{(r)}$ for $\beta=0$.\footnote[3]{We thank a referee for clarifying the argument required here.}
%




To find the precise form of the potential
$V_{\ell}{(r)}$ for $\ell=0$ and $\beta =0$, we impose the
requirement that there is no charge at the center of the star
$r=0$. To examine the charge at the center of the star at $r=0$,
we consider a small spherical Gaussian surface at $r=\epsilon$
around the center. When we integrate the field due to this
potential over this small Gaussian surface, 
the charge inside it is proportional to (see appendix \ref{gauss})
\[
\frac{r^{2}}{\beta^{2}+\alpha r^{2}} \frac{dV_{0}(r)}{dr}
\bigg|_{r=\epsilon}.
\]
Only the $\ell=0$ term would survive because the integral would
involve a term of $P_{\ell}{(\mathrm{cos\theta})}$ integrated over
the surface.  Requirement that the charge at the center should
vanish implies that for $\beta=0$, the potential $V_{0}{(r)}$
takes the form of $h_{0}{(r)}$ which is a constant. There is no
ambiguity.  






Since we already have the singular field $V^{\rm S}$ outside the
surface of the star from (\ref{8}) and (\ref{9}), we can now
extend it to the interior of the star.  Continuity of the vector
potential $A_{\mu}$ at the surface of the
star along with the coordinate transformation that takes the
interior star metric from (\ref{2}) to (\ref{3}) implies that
\begin{equation}
A_{t}|_{r=r_{s}}=\frac{\partial T}{\partial t}A_{T}|_{R=R_{s}}.
\end{equation}
From (\ref{3b}), we have ${\partial T}/{\partial t}={4}/{(2a+1)}$.
Putting everything together, we see that, for $\beta\neq0$, we can
write the singular potential $V^{\rm S}$ inside the star as
\begin{equation}
V_{\ell}^{\rm S}(r)=\frac{4e}{(2a+1)}\left\{
\begin{array}{cc}
f_{\ell}(b)g_{\ell}(R_{s})\frac{h_{\ell}(r)}{h_{\ell}(r_{s})} &\text{for } \ell\neq0  \\
\vbox{\vskip0.2truein}
f_{0}(b)g_{0}(R_{s})-\frac{M}{R_{s}b} &\text{ for  } \ell=0.\\
\end{array}
\right.
\label{14}
\end{equation}
For $\beta=0$, we have $a=1/2$, which gives ${\partial
T}/{\partial t}=2$, in which case
\begin{equation}
\label{14b}
%
V_{\ell}^{\rm S}(r)=2e\left\{
\begin{array}{cc}
f_{\ell}(b)g_{\ell}(R_{s})(\frac{r_{s}}{r})^{\ell} &\text{for  } \ell\neq0 \\
\vbox{\vskip0.2truein}
f_{0}(b)g_{0}(R_{s})-\frac{M}{R_{s}b} &\text{ for  } \ell=0.
\end{array}
\right.
\end{equation}


\section{Self force of the charge}\label{regular}

The static potential we have just constructed has a discontinuous derivative at the surface of the star, implying a surface charge density, $\sigma$, there.  In this section, we shall deal with $\sigma$ in two distinct ways, through a potential $\widetilde{V}$, that will actually be responsible for the self-force, which we evaluate
in the two specific cases we consider:
\begin{itemize}
\item{when the star is insulating, cannot be polarized, and has no free charges whatsoever; continuity of the gradient of the potential across the surface will ensure no charge is required there,}
\item{when the
star is highly conducting, with zero net charge; 
free charges arrange themselves on the surface so that there is no electric field
inside the star.}
\end{itemize}
In each case, we obtain the regular potential
$V^{\rm R}$, by imposing suitable boundary conditions on the surface of
the star.

\subsection{Insulating Star}\label{insulate}

The singular field $V^{\rm S}$ constructed in section \ref{staticEM} demands
the existence of a charge density on the surface of the star. Knowing
the field outside the star from (\ref{9}) and inside the
star from (\ref{14}), we can apply Gauss's law (see 
appendix \ref{gauss}) to find  this charge density.  
Writing the area element on the Gaussian surface $\sin\theta
d\theta d\phi,$ as $d\Omega$, and following (\ref{A2}) from the
appendix on Gauss's law, we obtain:
\begin{equation}
4\pi\sigma(\theta) d\Omega =-\left.R^{2}\frac{\partial V^{\rm
S}(R,\theta)}{\partial R}\right|_{R=R_{s}}d\Omega
+\left.\frac{r^{2}}{\beta^{2}+\alpha r^{2}}\frac{\partial V^{\rm
S}(r,\theta)}{\partial r}\right|_{r=r_{s}}d\Omega .\label{16}
\end{equation}
For the extreme density star
($\beta=0$), we can re-express (\ref{16}) in terms of the
multipole moments of the charge density and the potential using equations (\ref{9}) and (\ref{14b}):

\begin{equation}
4\pi\sigma^{\ell}=-e\left\{
\begin{array} {cc}
R_{s}^{2}f_{\ell}(b){g_{\ell}}'(R_{s})+ \frac{2\ell}{\alpha r_{s}}f_{\ell}(b)g_{\ell}(R_{s}) &\text{for }   \ell\neq0 \\
\vbox{\vskip0.2truein}
\frac{M}{b} &\text{ for  }  \ell=0. \label{17}
\end{array}
\right.
\end{equation}
Here, $\sigma^{0}$ corresponds to the net charge on the surface of
the star. It is clear that, in general, higher  moments of the charge
distribution are also non vanishing.

Since an insulating star should not have any charge distribution on
its surface, we have to add a regular potential $V^{\rm R}$ to the
singular potential $V^{\rm S}$, so that the full potential satisfies
the boundary condition of zero charge on the surface. Clearly,
$V^{\rm R}$ is the (otherwise homogeneous) potential obtained by a charge density of
$-\sigma(\theta)$ on the stellar surface. We call this potential
$\widetilde{V}$. As usual, we shall decompose this potential into
multipole moments:
$\widetilde{V}(R,\theta)=\Sigma_{\ell}\widetilde{V}_{\ell}(R)P_{\ell}(\cos\theta)$.

As we already know from \cite{wald}, the two linearly independent
vacuum solutions for $\widetilde{V}_{\ell}(R)$ outside the star's
surface are $f_{\ell}(R)$ and $g_{\ell}(R)$ from (\ref{11}) and (\ref{10}).
For large $R$, $f_{\ell}(R)\rightarrow R^{-(\ell+1)}$ and
$g_{\ell}(R)\rightarrow R^{\ell}$. Since we require
$\widetilde{V}(R,\theta)\rightarrow0$ for large $R$, we need
$\widetilde{V}_{\ell}(R)=C_{\ell}f_{\ell}(R)$.  The $C_{\ell}$ are constant
coefficients whose values are yet to be determined. Before finding
these coefficients, we must first find the form of the
potential $\widetilde{V}_{\ell}(R)$ inside the star.

By requiring $\widetilde{V}_{\ell}(r)$ inside the surface of the star to impose no source at $r\!=\!0$, we can fix its form everywhere inside the star.
For the extreme density star ($\beta =0$), we require
$\widetilde{V}_{\ell}(r)=D_{\ell}/r^{\ell}$
for all $\ell$, 
where the $D_{\ell}$ are 
constant
coefficients. Imposing the continuity of $\widetilde{V}_{\ell}$ across
the star's surface, along with the coordinate transformation which
takes the metric from (\ref{2}) to (\ref{3}), gives a relation
between the $C_{\ell}$ and $D_{\ell}$ coefficients:
\begin{equation}
D_{\ell}=\frac{\partial T}{\partial
t}C_{\ell}\,f_{\ell}(R_{s})r_{s}^{\ell},\qquad {\rm and} \qquad
D_{0}=\frac{\partial T}{\partial t}C_{0}{1\over R_{s}}.
\end{equation}
Since the charge on the surface is $-\sigma(\theta)$, we can use
Gauss's law (as shown in the appendix) to write an equation
analogous to (\ref{16}):
\begin{equation}
-4\pi\sigma(\theta) d\Omega
=-\left.R^{2}
\frac{\partial\widetilde{V}(R,\theta)}{\partial
R}\right|_{R=R_{s}} d\Omega
+\left.\frac{r^{2}}{\beta^{2}+\alpha
r^{2}}
\frac{\partial\widetilde{V}(r,\theta)}{\partial
r}\right|_{r=r_{s}} d\Omega
.\label{19}
\end{equation}
For the extreme density case ($\beta=0$), we can re-express
(\ref{19}) in terms of the multipole moments.
\begin{equation}
-4\pi\sigma^{\ell}=\left\{
\begin{array}{cc}
-R_{s}^{2}C_{\ell}{f_{\ell}}'(R_{s})-2\frac{2\ell}{\alpha r_{s}}C_{\ell}f_{\ell}(R_{s}) &\text{for  } \ell\neq0 \\
\vbox{\vskip0.2truein}
C_{0} &\text{ for  } \ell=0.
\end{array}
\right.
\label{20}
\end{equation}
Now (\ref{17}) and (\ref{20}) can be solved for the coefficients $C_{\ell}$.
For $\ell\!=\!0$, we simply have $C_{0}\!=\!eM/b$. For $\ell\!\neq\!0$, we find:
\begin{equation}
%
 -\frac{C_{\ell}}{e}\left[R_{s}^{2}{f_{\ell}}'(R_{s})+\frac{2\ell}{\alpha
r_{s}}f_{\ell}(R_{s})\right]=R_{s}^{2}
f_{\ell}(b){g_{\ell}}'(R_{s})
+\frac{2\ell}{\alpha r_{s}}
f_{\ell}(b)g_{\ell}(R_{s})
.
\label{21}
\end{equation}

\subsubsection{Evaluating the self force}

The regular potential in the vicinity of the charge has been
completely determined: $\widetilde{V}_{\ell}(R)=C_{\ell}f_{\ell}(R)$. The self
force experienced  by the charge $e$ is now given simply by Lorentz
force law $\overrightarrow{F}=e\overrightarrow{E}$, where
$\overrightarrow{E}$ is the electric field measured by a locally
inertial observer at rest on the charge $(R=b,\theta=0)$. Since the
electromagnetic stress tensor is a rank 2 tensor, under a coordinate transformation,
$F_{\alpha\beta}=F_{\mu\nu}(\partial x^{\mu}/\partial
x^{\alpha})(\partial x^{\nu}/\partial x^{\beta})$, where we take $(\mu,\nu)$ to
correspond to Schwarzschild coordinates and $(\alpha,\beta)$ to
correspond to the locally inertial coordinates,
which we denote by $\{{\cal R,T},\Theta,\Phi\}$. The diagonal form
of metric in Schwarzschild coordinates, implies a simple coordinate
transformation to the locally inertial coordinates, $\partial
x^{\mu}/\partial
x^{\alpha}=\delta^{\mu}_{\alpha}(1/\sqrt{\left|g_{\mu\mu}\right|})$.
The electric field in the locally inertial coordinate system is then
$E_{\cal I}=
F_{0i}/\sqrt{\left|g_{00}\right|g_{ii}}$.

\begin{equation}
E_{\cal R}=-\frac{\partial\widetilde{V}(R,\theta)}{\partial R}\,;\qquad
E_{\Theta}=-\frac{1}{R}\frac{1}{\sqrt{1-2M/R}}\frac{\partial\widetilde{V}(R,\theta)}{\partial\theta}.\label{22}
\end{equation}
On the z-axis ($\theta=0$), we have $E_{\theta}=0$, because, for all
$\ell$,  $\left.{\partial
P_{\ell}(\cos\theta)}/{\partial\theta}\right|_{\theta=0}=0.$
 The force experienced by the charge is therefore in the radial direction.
Since $\left.P_{\ell}(\cos\theta)\right|_{\theta=0}=1$, the self force
can be summed up over different components of $\ell$, to give
\begin{equation}
F_{\mathrm{self}}=\sum_{\ell}F_{\mathrm{self}}^{\ell}=-e\sum_{\ell}\left.C_{\ell}\frac{\partial
f_{\ell}(R)}{\partial R}\right|_{R=b}.\label{23}
\end{equation}
 Using (\ref{21}) to substitute for $C_{\ell}$, we get
\begin{equation}
 F_{\mathrm{self}}=-\frac{e^{2}M}{b}\,
{f_{0}}'(b)+e^{2}\sum_{\ell=1}^{\infty}\left(\frac{R_{s}^{2}f_{\ell}(b){g_{\ell}}'(R_{s})+(2\ell)f_{\ell}(b)g_{\ell}(R_{s})/(\alpha
r_{s})}{R_{s}^{2}{f_{\ell}}'(R_{s})+(2\ell)f_{\ell}(R_{s})/(\alpha
r_{s})}\right){f_{\ell}}'(b).
\label{24}
\end{equation}
%
%

\begin{figure}[htbp]
\scalebox{0.45}{ \psfig{file=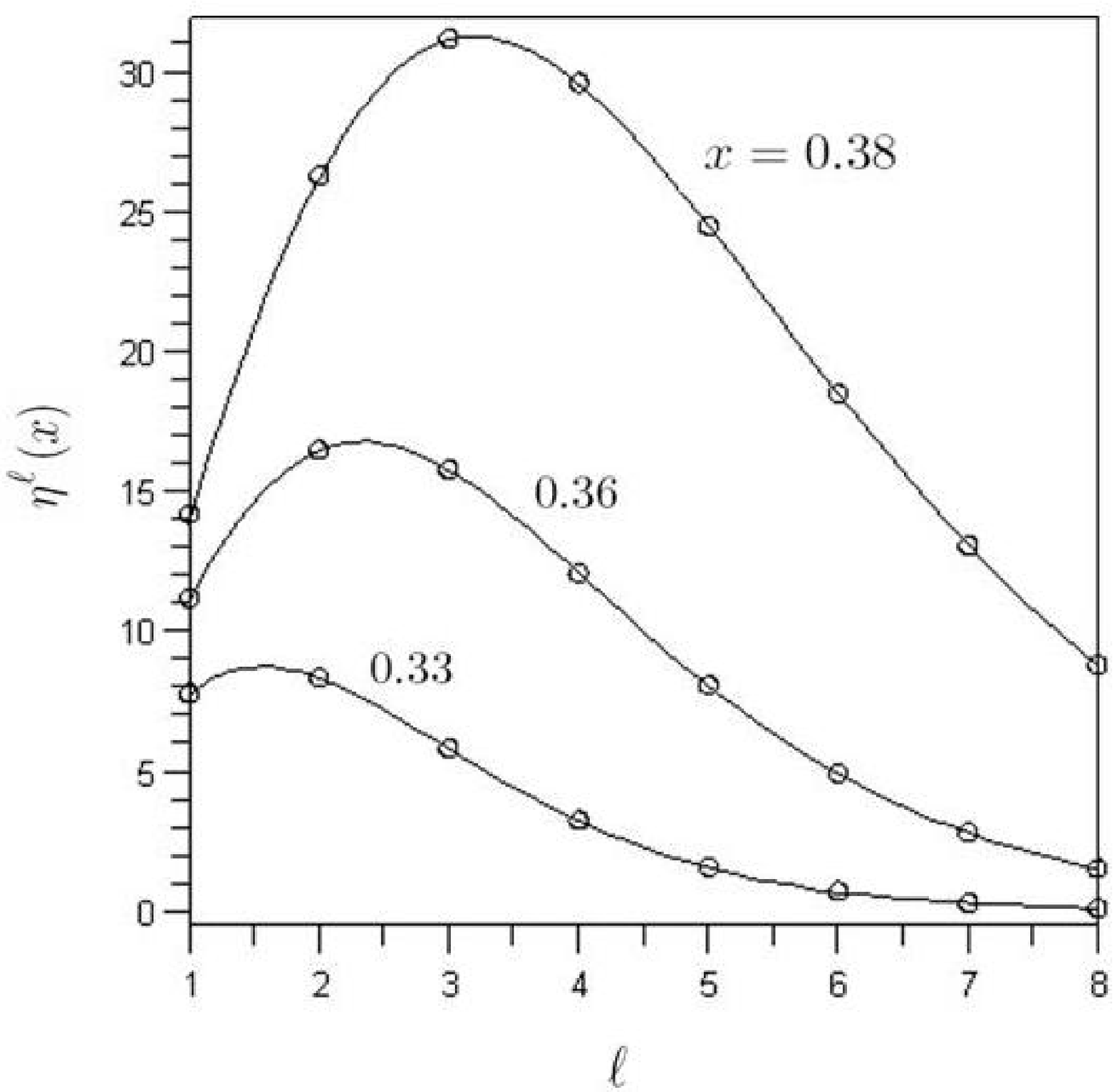} }~~\,~~\scalebox{0.45}{
\psfig{file=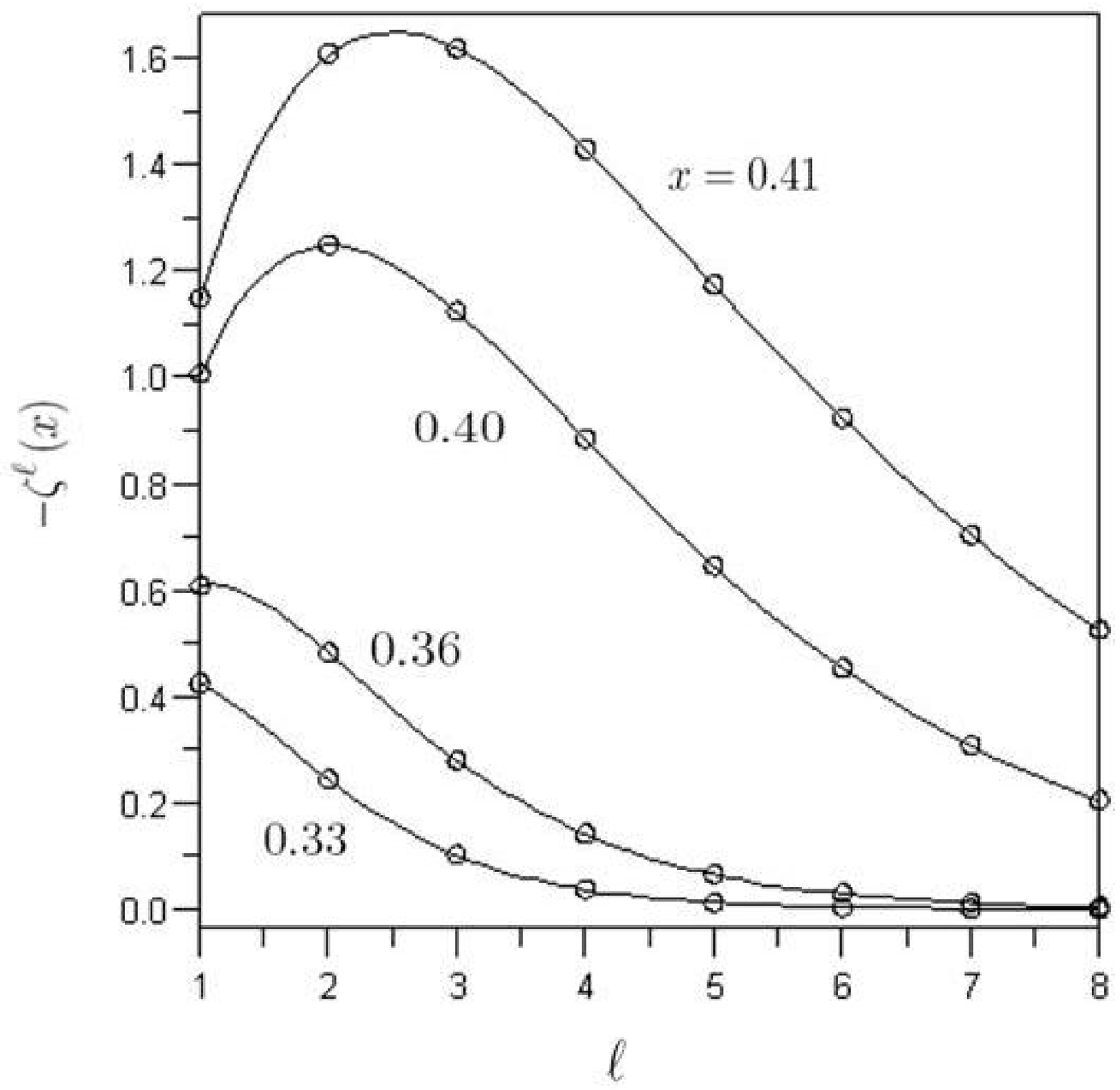} }
\caption{For various values of x, the convergence 
 of $\eta^{\ell}{\rm(x)}$ and $\zeta^{\ell}{\rm(x)}$ in $\ell$ is shown.  Note the difference in vertical scales.}\label{fig1}
\end{figure}

With a Schwarzschild black hole, as
in \cite{Will}, in place of the star, the self force experienced by the charge is
just
$F_{\mathrm{self}}^{\mathrm{BH}}={e^{2}}/{b^{2}}\left[{M}/{b}\right]$.
We now explicitly compute  $F_{\mathrm{self}}$ from (\ref{24})
and compare it with $F_{\mathrm{self}}^{\mathrm{BH}}$. From
section \ref{metric}, we know that for the extreme density star,
$R_{s}={9M}/{4}$, $r_{s}={27M}/{16}$ and $\alpha r_{s}={8}/{27M}$.
We can expand the functions $f_{\ell}$ and $g_{\ell}$ in (\ref{24}) to
get an explicit expression for $F_{\mathrm{self}}^{\ell}$. 
Each $F_{\mathrm{self}}^{\ell}$ can be written as
\begin{equation}
F_{\mathrm{self}}^{\ell}=\frac{e^{2}}{b^{2}}\left[\frac{M}{b}\right]\eta^{\ell}\left(\frac{M}{b}\right),\label{281}\end{equation}
where we have factored out ${e^{2}}/{b^{2}}\left[{M}/{b}\right]$ from each
$F_{\mathrm{self}}^{\ell}$, to show its relationship with
$F_{\mathrm{self}}^{\mathrm{BH}}$.
For convenience, we define $x\!=\!{M}/{b}$, and give the exact form of $\eta^{\ell}(x)$ for $\ell\!=\!0,1,2$:
\begin{equation}
\eta^{0}(x)=1,\label{25}
\end{equation}
\begin{eqnarray}
\eta^{1}(x) & = & 
\frac{81}{2x^{4}(27\ln3-28)}\left(\left[x-\frac{1}{2}\right]\ln\left[1-2x\right]-x+x^{2}\right)
\nonumber \\
 &  & \times
 \left(\frac{1}{2}\ln\left[1-2x\right]+x+x^{2}\right)
 ,\label{26}
\end{eqnarray}
%




\begin{eqnarray}
\eta^{2}(x) & = & 
\frac{945}{2x^{6}(567\ln3-620)}\left(\left[\frac{9}{2}x-3\right]\ln\left[1-2x\right]+x^{3}+3x^{2}-6x\right)
\nonumber \\
 &  & \times
 \left(\left[3x^{2}-\frac{9}{2}x+\frac{3}{2}\right]\ln\left[1-2x\right]+x^{3}-6x^{2}+3x\right)
 .\label{27}
\end{eqnarray}
%
%
Note that  $x$ can take
values in the range $(0,{4}/{9})$, where $x=0$ corresponds to the charge
being at infinity and $x={4}/{9}$ corresponds to the charge being on the surface
of the star. It turns out that, for every $\ell$, $\eta^{\ell}(x)$ is
positive and for large $\ell$, $\eta^{\ell}(x)\rightarrow0$, for all $x$ in the specified range.
In Fig \ref{fig1}, the convergence
of $\eta^{\ell}(x)$ is plotted for various values of $x$. 
When the charge is close to the surface of the star ($x$ close to
$4/9$), we must sum over many $\ell$'s to get an accurate result for
the self force.

As an example, if the charge is placed at
$b=3M\Rightarrow(x={1}/{3})$, we can numerically calculate
$F_{\mathrm{self}}$ to be
\begin{equation}
F_{\mathrm{self}}=\frac{e^{2}}{b^{2}}\left[\frac{M}{b}\right]\Sigma_{\ell}\eta^{\ell}(\frac{1}{3})=31.8\,\frac{e^{2}}{b^{2}}\left[\frac{M}{b}\right].
\end{equation}
The self force experienced  by the charge at $b=3M$, in the
presence of this star is 31.8 times stronger than the force it
would experience if the star were replaced by an equally massive
black hole. As $x\rightarrow0$ $(b\rightarrow\infty)$, we have
$F_{\mathrm{self}}\rightarrow F_{\mathrm{self}}^{\mathrm{BH}}$:  
a charge placed far away from this star, would feel the same force
as it would feel when the star is replaced by a black hole of
equal mass. But as we move the charge close to the star
$x\rightarrow4/9$, $F_{\mathrm{self}}$ becomes orders of magnitude
greater than $F_{\mathrm{self}}^{\mathrm{BH}}$, for example, at
$x=0.43$, we have $F_{\mathrm{self}}=36777\,
F_{\mathrm{self}}^{\mathrm{BH}}$.

\subsection{Conducting Star}\label{cond}

So far, we have considered an insulating star which cannot be
polarized. Now, we turn our attention to a conducting star which
has free charges on it, but no net charge. Hence there could be non
zero induced surface charge density. We call this surface charge
density $\widetilde{\sigma}(\theta)$ (which can be decomposed as
$\widetilde\sigma(\theta)=\Sigma_{\ell}\left[\widetilde\sigma^{\ell}P_{\ell}(\cos\theta)\right]$).
This star would be polarized in such a way that the field inside it
vanishes. The interior metric of the star plays no role in
determining the field outside the star, only the size of the star
$R_{s}$ matters. As before, $V=V^{\rm S}+\widetilde{V}$ is the total
potential, $V^{\rm S}$ is the singular piece which exerts no self
force and $\widetilde{V}$ (homogenous in the vicinity of the particle) is
obtained by incorporating appropriate boundary conditions (namely
zero field in the interior of the star, and zero net charge on its surface).
Since the electric field is zero inside the star, the potential
$V^{\rm S}+\widetilde{V}$ should be a  constant on the surface of
the star, hence $V_{\ell}^{\rm S}(R_{s})+\widetilde{V}_{\ell}(R_{s})=0$
for $\ell\neq0$. For reasons already stated in section \ref{insulate},
$\widetilde{V}_{\ell}(R)=C_{\ell}f_{\ell}(R)$ outside the surface of the
star. Using (\ref{9}) for $V^{S}_{\ell}$, we can write
\begin{equation}
e\,g_{\ell}(R_{s})f_{\ell}(b)+C_{\ell}f_{\ell}(R_{s})=0:\qquad \ell\neq0.\label{28}
\end{equation}
This equation can be solved to obtain $C_{\ell}$ for all $\ell\neq 0$.

$C_{0}$  can be obtained by imposing the condition that the net
charge on the star vanishes, that is $\widetilde{\sigma}^{0}=0$. To
impose this condition, we should first apply Gauss's law to obtain
an expression for the induced surface charges
$\widetilde{\sigma}(\theta)$. Since the electric field inside the star
vanishes, applying Gauss's Law simply gives
(see appendix \ref{gauss}):
\begin{equation}
4\pi\widetilde{\sigma}(\theta)
=-\left.R^{2}
\frac{\partial V^{\rm S}(R,\theta)}{\partial
R}\right|_{R=R_{s}}
-\left.R^{2}
\frac{\partial\widetilde{V}(R,\theta)}{\partial
R}\right|_{R=R_{s}}
\end{equation}
\begin{equation}
\Rightarrow4\pi\widetilde{\sigma}^{\ell}=\left\{
\begin{array}{cc}
-R_{s}^{2}\left[e\,f_{\ell}(b){g_{\ell}}'(R_{s})+C_{\ell}{f_{\ell}}'(R_{s})\right]\ne0 &\text{for  } \ell\neq0 \\
\vbox{\vskip0.2truein}
-R_{s}^{2}\left[eM/R_{s}^{2}b -C_{0}/R_{s}^{2}\right] &\text{ for  } \ell=0.\\
\end{array}
\right.
\end{equation}
In particular, using the fact that $\widetilde{\sigma}^{0}=0$, we find that
$C_{0}=eM/b$.  Note that $V_{0}(r)=2e/b$ for $r<r_{s}$.

\subsubsection{Evaluating the self force}

Now that we know all the coefficients $C_{\ell}$, we have completely
determined the regular field that would produce the self force. As
in the previous section (insulating star), (\ref{23}) gives the
self force on the charge, with the only change being that the
$C_{\ell}$ are now given by (\ref{28}) rather than (\ref{21}). So, the
self force is
\begin{equation}
F_{\mathrm{self}}=\sum_{\ell}F_{\mathrm{self}}^{\ell}=-\frac{e^{2}M}{b}\,
{f_{0}}'(b)+e^{2}\sum_{\ell=1}^{\infty}\left(\frac{f_{\ell}(b)g_{\ell}(R_{s})}{f_{\ell}(R_{s})}\right){f_{\ell}}'(b).\label{30}
\end{equation}

\begin{figure}
\begin{center}
\includegraphics[
scale=0.70]
  {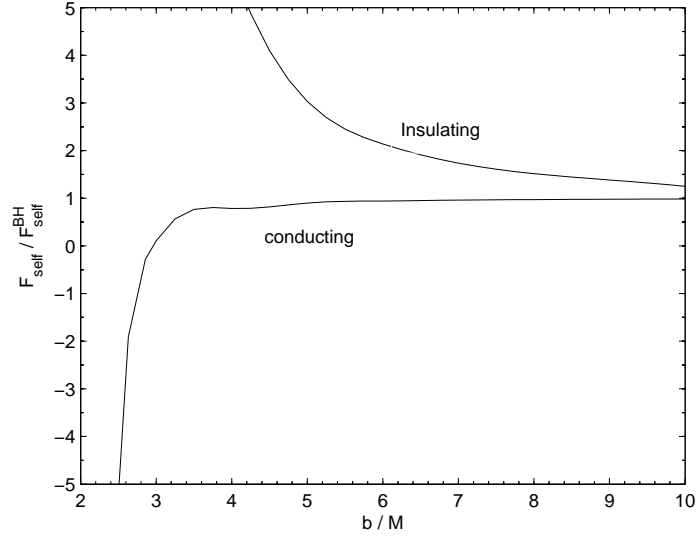}
\caption{The self-force on the charge is plotted as a function of its position for the two cases considered in the text:  when the star is an insulator (upper curve) and when it is a conductor (lower curve).}\label{fig2}
\end{center}
\end{figure}

\noindent Observe that the result obtained for the self-force in (\ref{30})
does not depend on the interior metric of the star.

To compare the results with that of the insulating star, we assume
that the size of the star $R_{s}$ is the same as that of an
extreme density star, $R_{s}={9M}/{4}$. Once again, it turns out
that each $F_{\mathrm{self}}^{\ell}$ can be expressed as
\begin{equation}
F_{\mathrm{self}}^{\ell}=\frac{e^{2}}{b^{2}}\left[\frac{M}{b}\right]\zeta^{\ell}\left(\frac{M}{b}\right).\label{31}\end{equation}
We again define $x={M}/{b}$ for convenience. The form of
$\zeta^{\ell}(x)$ is given for $\ell=0,1,2$.
\begin{equation}
\zeta^{0}(x)=1,\label{32}
\end{equation}
\begin{eqnarray}
\zeta^{1}(x) & = & -
\frac{108}{x^{4}(72\ln3-160)}\left(\frac{1}{2}\ln\left[1-2x\right]+x+x^{2}\right)
\nonumber \\
 &  & \times
 \left(-\left[\frac{1}{2}-x\right]\ln\left[1-2x\right]-x+x^{2}\right)
 ,\label{33}
\end{eqnarray}
\begin{eqnarray}
\zeta^{2}(x) & = & -
\frac{900}{x^{6}(1080\ln3-1376)}\left(\left[\frac{9}{2}x-3\right]\ln\left[1-2x\right]+x^{3}+3x^{2}-6x)\right)
\nonumber \\
 &  & \times
 \left(\left[\frac{3}{2}-\frac{9}{2}x+3x^{2}\right]\ln\left[1-2x\right]-6x^{2}+x^{3}+3x\right)
 .\label{34}
\end{eqnarray}

Whereas the $\eta^{\ell}(x)$ are positive for all
values of $\ell$, the $\zeta^{\ell}(x)$ are negative for all $\ell$ except
$\ell=0$. For large $\ell$, $\zeta^{\ell}(x)\rightarrow0$, for all $x$.
The convergence of $\zeta^{\ell}(x)$ is also plotted in Fig \ref{fig1} for various values of $x$.
We again note that as $x\rightarrow0$ $(b\rightarrow\infty)$, we
have $F_{\mathrm{self}}\rightarrow F_{\mathrm{self}}^{\mathrm{BH}}$;
a charge placed far away from this star, would feel the
same force as it would feel if the star were replaced by an equally
massive black hole. However, something interesting happens nearby.
There exists a point, $x_{0}$, where the charge feels zero
self-force. This is an electrostatic equilibrium point. It happens
when $\sum_{\ell=1}^{\infty}\zeta^{\ell}(x_{0})=-\zeta^{0}$. This condition can be
numerically solved, giving $x_{0}=0.34$ $(b=2.95M)$.  When $x<x_{0}$($b>2.95M$), the charge would experience a repulsive
force and when $x>x_{0}$($b<2.95M$), the charge would experience an
attractive force.

As an example, we shall numerically compute the
force experienced by the  charge  placed at $x=0.42$.
\begin{equation}
F_{\mathrm{self}}=\frac{e^{2}}{b^{2}}\left[\frac{M}{b}\right]\Sigma_{\ell}\zeta^{\ell}(0.42)=-19.96\frac{e^{2}}{b^{2}}\left[\frac{M}{b}\right].
\end{equation}
The closer we take the charge to the surface of the star, the more the
attractive force increases. In the limit $x\rightarrow4/9$,
$F_{\mathrm{self}}\rightarrow-\infty$. This is expected because even
in flat space, when a charge is placed close to a conductor, it
experiences a huge attractive force. We should interpret the
attractive force which dominates when the charge is close to the
conducting star as a consequence of electrical polarization of the
star and the repulsive force which dominates when the charge is far
away from the star as due to interaction of the fields with
space-time curvature.
 Fig \ref{fig2} presents the self force
experienced by the charge $F_{\mathrm{self}}$ (normalized with
respect to $F_{\mathrm{self}}^{\mathrm{BH}}$) as a function of the
position of the charge placed near a conducting and an insulating
star.


%

A simple consistency check can be performed in the limiting case
when $M\rightarrow 0$, by comparing it to the well known flat
space limit. In flat space, the self force on the charge $e$ at a
distance $b$ from the center of a conducting sphere of radius
$R_{s}$ and zero net charge can be calculated to be

\begin{equation}
F_{\mathrm{self}}^{\mathrm{flat}}=\frac{e^{2}}{b^{2}}\left[\frac{1}{\left(1-\frac{R_{s}^{2}}{b^{2}}\right)^{2}}-1\right]
\left(\frac{R_{s}}{b}\right). \label{FS}
\end{equation}
In the limit $M\rightarrow 0$, (\ref{11},\ref{10}) reduce to
$f_{\ell}(R)\rightarrow 1/R^{\ell +1}$ and $g_{\ell}(R)
\rightarrow R^{l}$. As a result (\ref{30}) reduces to

\[
F_{\mathrm{self}}=\frac{e^{2}}{b^{2}}\sum_{\ell=1}^{\infty}(\ell
+1)\left(\frac{R_{s}}{b}\right)^{2\ell +1},
\]
which exactly matches the flat space result (\ref{FS}).
%

\appendix

\section{Gauss's Law}\label{gauss}

In this appendix, we 
apply Gauss's law to evaluate the 
charge density on the star's surface. We 
start by reviewing 
Gauss's law. Consider  a spacelike 3-hypersurface $\Sigma$, bounded
by a 2-surface $\partial\Sigma$. Then Gauss (Strokes) theorem can be
written as \cite{poisson}
\begin{equation}
\int_{\Sigma}\nabla_{\nu}F^{\mu\nu}\,
d\Sigma_{\mu}=\frac{1}{2}\oint_{\partial\Sigma}F^{\mu\nu}dS_{\mu\nu}.\label{A1}\end{equation}
From Maxwell's equations, the LHS corresponds to $-4\pi$ times $Q$, the total
charge enclosed within the bounded region $\Sigma$. The surface
element $dS_{\mu\nu}=-2n_{[\mu}\gamma_{\nu]}\sqrt{\sigma}d\theta
d\phi$, where $n_{\mu}$ is the future directed unit normal to the
space like hypersurface $\Sigma$ and $\gamma_{\nu}$ is the unit
normal to the  boundary $\partial\Sigma$. The vector $\gamma_{\nu}$
should always be directed away from the surface $\partial\Sigma$.
The term $\sqrt{\sigma}d\theta d\phi$, represents the 2 dimensional
surface area element.


In this paper, since we are interested in the surface charge
density of the star, we 
apply 
Gauss's law in a region
very close the surface of the star. 
We choose the
hypersurface $\Sigma$ to be a constant time $T$ (or $t$)
hypersurface spatially bounded by the Gaussian surface $\partial\Sigma$,
which is made up of two small pieces of constant-$R$ surfaces of
coordinate area $d\theta d\phi$, one just outside the star surface
$R=R_{s}$ (\ref{1}) and the other just inside the star surface $r=r_{s}$ (\ref{3a}).

For the piece of $\partial\Sigma$ outside the star's surface, in the
Schwarzschild coordinate system $(T,R,\theta,\phi)$ as in metric
(\ref{2}), $n_{\mu}=(-\sqrt{1-2M/R},0,0,0)$ and
$\gamma_{\mu}=(0,{1}/{\sqrt{1-2M/R}},0,0)$. The only nonvanishing
component of $dS_{\mu\nu}$ is $dS_{TR}=R^{2}\sin\theta d\theta
d\phi$. For the piece of $\partial\Sigma$ inside the surface of the
star, $dS_{\mu\nu}$ is obtained from the unit normals to
constant-$t$ and constant-$r$ surfaces with respect to the star's
interior metric. With respect to the star's coordinate system
$(t,r,\theta,\phi)$ as in metric (\ref{3a}),
$n_{\mu}=(-(\beta^{2}+\alpha r^{2}),0,0,0)$ and
$\gamma_{\mu}=(0,-1,0,0)$. Here, the only nonvanishing component of
$dS_{\mu\nu}$ is $dS_{tr}=-(\beta^{2}+\alpha r^{2})r^{2}\sin\theta
d\theta d\phi$.

By applying Gauss's Law, (\ref{A1}), on $\Sigma$, we obtain,
\medskip
\begin{equation}
-4\pi\sigma(\theta,\phi)\sin\theta d\theta d\phi =
\left.g^{TT}g^{RR}F_{TR}dS_{TR}\right|_{\mathrm{R=R_{s}}}^{\mathrm{outside}}+\left.g^{tt}g^{rr}F_{tr}dS_{tr}\right|_{\mathrm{r=r_{s}}}^{\mathrm{inside}}.\label{A2}
\end{equation}
\medskip
\noindent Here, $\sigma(\theta,\phi)$ is the surface charge density and
$\sigma(\theta,\phi)\sin\theta d\theta d\phi$ is the total charge
contained within that small Gaussian surface. The axial symmetry of
the problem considered in this paper ensures that
$\sigma(\theta,\phi)$ is actually just $\sigma(\theta)$.
We have  adopted (\ref{A2}) in section \ref{regular} of this paper. For the
case of conducting star (section \ref{cond}), we note that the field
inside the star vanishes and hence the second term in the RHS of
(\ref{A2}) vanishes.

\subsection{Conformal invariance of Maxwell's equations}

As already indicated, we can ignore the conformal factor in (\ref{3}) and use the metric  (\ref{3a}) when
evaluating the integral $\frac{1}{2}\int F^{\mu\nu}dS_{\mu\nu}$ on
the surface just inside the star. This is justified 
as long as the source
($J^{\mu}\sqrt{-g}$) is conformally invariant, in which case: 
\begin{itemize}
\item
the LHS of (\ref{A1}) (that is, the total charge contained within a coordinate volume) is invariant
under a conformal transformation, 
\item
Maxwell's equations have the
same solution for the vector potential $A_{\mu}$, and hence the same
solution for $F_{\mu\nu}$, in any conformally related metric.
\end{itemize}
Note that, under a conformal transformation
$g_{\mu\nu}\!\rightarrow\!\Omega^{2}g_{\mu\nu}$
($g^{\mu\nu}\!\rightarrow\!\Omega^{-2}g^{\mu\nu}$), so we also have
$n_{\mu}\!\rightarrow\!\Omega n_{\mu}$,
$\gamma_{\mu}\!\rightarrow\!\Omega\gamma_{\mu}$,
$\sqrt{-g}\!\rightarrow\!\Omega^{4}\sqrt{-g}$ and
$\sqrt{\sigma}\!\rightarrow\!\Omega^{2}\sqrt{\sigma}$. Hence,
\begin{equation}
dS_{\mu\nu}\rightarrow\Omega^{4}dS_{\mu\nu}. \label{A3}
\end{equation}
Also note that, if the source ($J^{\mu}\sqrt{-g}$) is conformally
invariant, Maxwell's equations (\ref{conf}) require  $F_{\mu\nu}$ to
be conformally invariant. With contravariant indices $F^{\mu\nu}$
transforms under a conformal transformation as
\begin{equation}
F^{\mu\nu}=g^{\alpha\mu}g^{\beta\nu}F_{\alpha\beta}\rightarrow\frac{g^{\alpha\mu}}{\Omega^{2}}\frac{g^{\beta\nu}}{\Omega^{2}}F_{\alpha\beta}=\frac{1}{\Omega^{4}}F^{\mu\nu}.\label{A4}
\end{equation}
Cleary, (\ref{A3}) and (\ref{A4}) imply that (\ref{A1}) is invariant under a conformal transformation.


\section*{References}
\begin{thebibliography}{10}
\bibitem{wald}J.M.~Cohen and R.M.~Wald, J.~Math.~Phys.~{\bf 12}, 1845 (1971).
\bibitem{hanni}R.S.~Hanni and R.M.~Ruffini, Phys.~Rev.~{\bf D8}, 3259 (1973).
\bibitem{copson}E.T.~Copson, Proc.~Roy.~Soc.~London {\bf A118}, 184 (1928).
\bibitem{linet}B.~Linet, J.~Phys.~{\bf A9}, 1081 (1976).
\bibitem{hadamard}J. Hadamard, {\it Lectures on Cauchy's Problem} (Yale University Press, New Haven, CT 1923).
\bibitem{dirac}P.A.M.~Dirac, Proc.~Roy.~Soc.~London {\bf A167}, 148
(1938).
\bibitem{Will}A.G.~Smith and C.M.~Will, Phys.~Rev.~{\bf D22}, 1276 (1980).
\bibitem{dewitt:brehme}B.S.~DeWitt and R.W.~Brehme, Ann. Phys. (N.Y.), {\bf 9}, 220-259 (1960).
\bibitem{bernard}S. Detweiler and  B.F.~Whiting, Phys.~Rev.~{\bf D67},
024025 (2003).
\bibitem{barack:ori}L.~Barack and A~ Ori, Phys.~Rev.~{\bf D61}, 061502 (2000).
\bibitem{quinn:wald}T.C.~Quinn andÊ R.M.~Wald, Phys.~Rev.~{\bf D56}, 3381 (1997).
\bibitem{burko:etal}L.M.~Burko, Y.T.~Liu and Y.~Soen, Phys.~Rev.~{\bf D63}, 024015 (2000).
\bibitem{dewitt}C.M.~DeWitt and B.S.~DeWitt, Physics (Long Island City, N.Y.) 1, 3 (1964). 
\bibitem{wiseman}A.G.~Wiseman, Phys.~Rev.~{\bf D61}, 084014 (2000).
\bibitem{MTW}C.W.~Misner, K.S.~Thorne and J.A.~Wheeler,
{\it Gravitation} (Freeman, San Francisco 1973).
\bibitem{conf:cords}K.~Shankar and B.F.~Whiting, gr-qc/0706.4324.
\bibitem{Arfken}G.~Arfken, {\it Mathematical methods for physicists}
3rd ed (Academic Press, San Diego 1985).
\bibitem{poisson}E.~Poisson, {\it A relativists toolkit: The mathematics of black-hole mechanics} (Cambridge University Press, Cambridge 2004).
\end{thebibliography}
\end{document}